# A spectrally bright wavelength-switchable vacuum ultraviolet source driven by quantum coherence in strong-field-ionized molecules

Yuexin Wan [1,3,#], Zhaoxiang Liu [1,3,#], Jinping Yao [1,†], Bo Xu [1,3], Jinming Chen [1,3,4], Fangbo Zhang [1,3], Zhihao Zhang [1,3,4], Lingling Qiao [1], and Ya Cheng [1,2,5,6,*]

[1]*State Key Laboratory of High Field Laser Physics and CAS Center for Excellence in Ultra-intense Laser Science, Shanghai Institute of Optics and Fine Mechanics(SIOM), Chinese Academy of Sciences(CAS), Shanghai 201800, China*
[2]*Collaborative Innovation Center of Light Manipulations and Applications, Shandong Normal University, Jinan 250358, China*
[3]*University of Chinese Academy of Sciences, Beijing 100049, China*
[4]*School of Physical Science and Technology, ShanghaiTech University, Shanghai 200031, China*
[5]*Collaborative Innovation Center of Extreme Optics, Shanxi University, Taiyuan, Shanxi 030006, China*
[6]*Shanghai Research Center for Quantum Sciences, Shanghai 201315, China*

(Date: XXX)

We report generation of spectrally bright vacuum ultraviolet (VUV) and deep UV (DUV) coherent radiations driven by quantum coherence in tunnel-ionized carbon monoxide (CO) molecules. Our technique allows us to switch between multiple wavelengths provided by the abundant energy levels of molecular ions. The DUV/VUV sources can have arbitrary polarization states by manipulating the pump laser polarization. The superior temporal and spectral properties of the developed source give rise to a broadband Raman comb in the DUV/VUV region.

Coherent vacuum ultraviolet (VUV, 100-200 nm) and deep UV (DUV, 200-300 nm) light sources have been widely applied in numerous studies, such as biochemical kinetics [1,2], surface structure analysis [3,4], precision measurements [5,6], atmospheric chemistry [7,8], and nanolithography [9]. Free-electron lasers have been built to provide wavelength-tunable short-wavelength light sources with high intensity, good coherence, short pulse duration as well as well-defined polarization [10,11]. However, these large facilities are expensive and typically of large footprint sizes, which limit the widespread applications. Recently, table-top DUV/VUV sources have been developed based on nonlinear optical crystals of $KBe_2BO_3F_2$ family. The shortest wavelength reported so far is 149.8 nm using this approach as determined by the absorption edge of the crystal [12]. In comparison with solids, gaseous media usually have a broader transparent window, a higher damage threshold as well as a lower dispersion, facilitating the extension of coherent radiation towards shorter wavelengths. Either four-wave mixing (FWM) in the gas chamber [13,14], or the dispersion wave in the gas-filled fiber [15,16], enables generation of wavelength-tunable DUV/VUV radiations.

It is also of vital importance to effectively control the parameters of the table-top DUV/VUV coherent light sources such as the wavelength, bandwidth, pulse duration and polarization. For instance, the polarization manipulation of DUV/VUV sources is a key prerequisite for studying the optical properties of chiral species [17] and magnetic textures [18,19], which cannot be readily achieved in the DUV/VUV sources generated using nonlinear crystals due to the constrains of the phase matching. In this Letter, we demonstrate a table-top DUV/VUV coherent light source with unique characteristics including narrow bandwidth, short pulse duration, switchable wavelength, and fully controllable polarization. The DUV/VUV coherent radiations are generated through the resonant excitation of $CO^+$ ions following tunnel ionization of CO molecules. The abundant vibrational energy levels of $CO^+$ ions allow us to select the wavelength of DUV/VUV radiations. The resonant interaction of laser fields with $CO^+$ ions establishes the electronic coherence, giving rise to the generation of picosecond DUV/VUV pulses with a low beam divergence and a decent photon flux. Last but not least, the combination of femtosecond impulsive excitation and picosecond DUV/VUV probe light can facilitate efficient Raman scattering as we have observed Raman combs with more than 100 sidebands in the DUV/VUV region. The DUV/VUV Raman combs feature a femtosecond pulse train in the time domain, indicating that the combs are phased locked.

This experiment was performed with a Ti:sapphire laser, which delivers 800 nm, 40 fs, ~6 mJ laser pulses. The laser was split into three beams. One beam with the energy of 3.5 mJ was used as near-infrared (NIR) pump laser. The second beam with the energy of 2 mJ was launched into a nonlinear crystal to produce a wavelength-tunable UV laser. NIR and UV pump lasers with the controllable delay were focused into CO gas using a $f$=60 cm lens to induce DUV coherent radiations. Unless otherwise specified, polarizations of two lasers were set parallel to each other. The residual NIR pump laser with the energy of 1.9 mJ and DUV coherent radiation were collimated, and then were focused into another $O_2$ gas chamber using a $f$=100 cm concave mirror to generate Raman comb. The spectra of DUV coherent radiations and Raman comb were recorded by a grating spectrometer. By introducing the third NIR beams, the temporal information of Raman comb was measured by the cross-correlation technique.

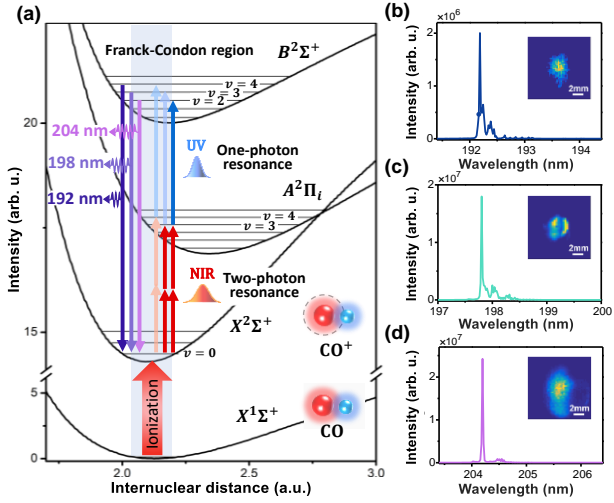

FIG. 1. (a) The potential energy curves of electronic states of $CO^+$ and neutral CO (Adapted from Ref. [21]). The Franck-Condon ionization region of CO molecules is indicated by the shadow. The red and blue arrows represent the two- and one-photon resonant excitations, respectively. Spectra of (b) 192 nm, (c) 198 nm, and (d) 204 nm radiations recorded at the delay of 1.26 ps and the gas pressure of 10 mbar. UV laser wavelength: (b), (c) 390 nm, (d) 415 nm. Inset: Beam profiles of the corresponding DUV/VUV radiations.

The electronic states of $CO^+$ ions produced by strong field ionization have suitable energy intervals and large dipole moments for generating coherent radiations in the DUV/VUV region [20]. As shown in Fig. 1(a), we choose three electronic states of $CO^+$ ions to construct a cascaded quantum system, and adopt NIR and UV pump lasers to realize resonant excitation among these energy states. The intensity of the NIR pump laser is sufficiently high to ionize CO molecules. The vibrational and rotational coherences are naturally created in $CO^+$ ions when the electron instantaneously escapes from its nucleus. Because the equilibrium internuclear distance of $X^1\Sigma^+$ state is comparable to that of $X^2\Sigma^+$ state, singly ionized CO molecules are mainly populated in $X^2\Sigma^+(v=0)$ state. $CO^+$ ions in this state can be efficiently excited to $A^2\Pi_i(v=3)$ and $A^2\Pi_i(v=4)$ states through resonant absorbing two NIR photons with the wavelength of 803.5 nm and 759.2 nm, respectively. Subsquently, $CO^+$ ions can be further excited to $B^2\Sigma^+(v=2,3,4)$ states by resonant absorbing one UV photon. The resonant excitation among three electronic states allows for strong DUV/VUV radiations by mixing frequency of NIR and UV laser pulses. Owing to fruitful vibrational energy levels involved in these electronic states, coherent radiations at 204 nm, 198 nm and 192 nm wavelengths can be selectively produced by tuning the wavelength of UV laser, which correspond to the transition from $B^2\Sigma^+(v=2,3,4)$ to $X^2\Sigma^+(v=0)$, respectively.

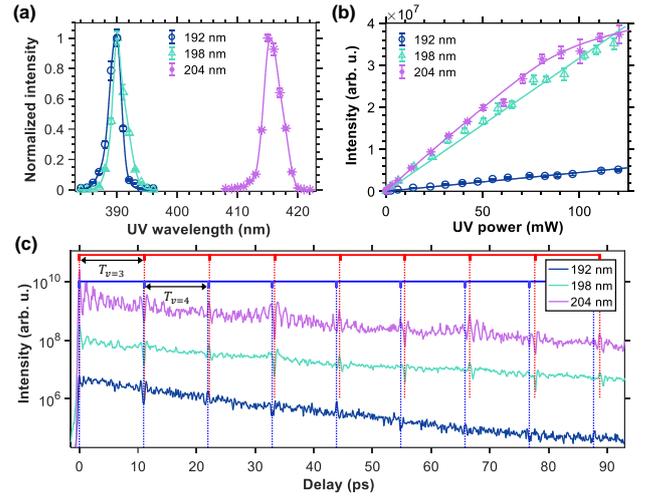

FIG. 2. (a) DUV/VUV radiations as a function of the UV laser wavelength. The power of UV laser is maintained at 45 mW while changing wavelength. The intensities of three radiations are normalized independently. (b) Dependence of DUV/VUV radiations on the UV laser power. The solid line is the least-squares fitting of experimental data. (c) Evolution of DUV/VUV signals with the time delay between NIR and UV pulses. The curves have been vertically shifted for clarity. The error bars in (a) and (b) represent the standard deviation of measurements.

The typical spectra of these DUV/VUV radiations are shown in Fig. 1(b)-(d). All signals exhibit a narrow bandwidth (~6 cm$^{-1}$), a small divergence angle (~4 mrad), a linear polarization, and a good spatial profile (see inset of Fig. 1(b)-(d)). In addition, these DUV/VUV radiations are overlapped with broadband FWM signals around zero delay, followed by a slow decay with the increasing delay of UV laser with respect to NIR laser. When UV laser pulses arrive ahead of NIR laser pulses, these narrow-bandwidth DUV/VUV radiations will vanish. Such a temporal evolution originates from the built up of electronic coherence during the resonant interaction of two pump lasers with $CO^+$ ions, which will be elaborated later.

We further studied the dependence of DUV/VUV coherent radiations on the wavelength and power of UV laser. Figure 2(a) clearly shows that 192 nm and 198 nm radiations reach the maximum when the UV laser wavelength is tuned to about 390 nm, whereas the 204 nm radiation is the strongest with the 415 nm UV laser. The optimal UV wavelengths for all radiations are in perfect agreement with that predicted by the resonant FWM scheme in Fig. 1(a). Furthermore, these radiations almost linearly grow with the increase of UV laser powers, as shown in Fig. 2(b). The intensity of the 204 nm signal is comparable to that of the 198 nm signal. Their maximum photon flux is estimated to about $4\times10^{10}$ photons/s. The maximum photon flux of the 192 nm signal is about $6\times10^9$ photons/s, which is weaker than other two signals due to the smaller Franck-Condon factor corresponding to this transition. In addition, the stronger spectral component at 803.5 nm is more beneficial to two-photon excitation from $X^2\Sigma^+(v=0)$ to $A^2\Pi_i(v=3)$ states and the subsequent 198 nm and 204 nm radiations.

We also examined the evolution of these DUV/VUV signals with the time delay between NIR and UV laser pulses. Unlike the conventional FWM which requires a temporal overlap between the interacting waves, the DUV/VUV radiations can still be generated when the NIR and UV pump laser pulses are temporally separated. As shown in Fig. 2(c), all the radiations last for tens of picoseconds. Meanwhile, the radiations decay slowly with an increasing delay between the two pump pulses, which reflects a long decoherence time between $X^2\Sigma^+$ and $A^2\Pi_i$ states of CO$^+$. In general, the depopulation is much slower than decoherence. Thus, we can estimate the decoherence time from the dynamics of these DUV/VUV radiations. In this experiment, the docoherence time between $X^2\Sigma^+(v=0)$ and $A^2\Pi_i(v=3)$ is measured to be about 30 ps, and the docoherence time between $X^2\Sigma^+(v=0)$ and $A^2\Pi_i(v=4)$ is about 16 ps. When scanning the NIR-UV delay, these radiations are strongly modulated at revival moments of rotational wavepackets of CO$^+$ ions. Interestingly, the modulation originates from rotational dynamics of intermediate state rather than upper and lower electronic states of these radiations. As shown in Fig. 2(c), modulation moments match very well with the revival periods of rotational wavepackets of $A^2\Pi_i(v=3,4)$ states, which are calculated by $T_{v=3,4}=1/(2B^*_{v=3,4}c)$ with the rotational constant $B^*_{v=3,4}$ of $A^2\Pi_i(v=3,4)$ state (i.e., $B^*_{v=3}$ =1.50 cm$^{-1}$ and $B^*_{v=4}$=1.52 cm$^{-1}$) [22,23]. This clearly indicates that the rotational coherence of CO$^+$ ions in the intermediate state is encoded in these coherent DUV/VUV radiations. Hence, such an experimental scheme for generation of a DUV/VUV coherent source also allows for exploring ultrafast dynamics of rotational wavepackets of excited states.

The polarization states of these DUV/VUV coherent radiations are fully controllable by changing the ellipticity of UV pump laser as we show below. For a resonant FWM process $\omega_{\text{FWM}}=2\omega_1+\omega_2$ in an isotropic nonlinear medium, the nonlinear polarization is given by [24,25]

$$\boldsymbol{P_R} = 3\chi^{(3)R}_{1111}(\omega_{\text{FWM}};\omega_1,\omega_1,\omega_2)\{(1-\bar{\rho})\boldsymbol{E_1}(\boldsymbol{E_1}\boldsymbol{E_2}) +\bar{\rho}\boldsymbol{E_2}(\boldsymbol{E_1}\boldsymbol{E_1})\}, \quad (1)$$

Where $\bar{\rho} = \chi^{(3)R}_{1221}/\chi^{(3)R}_{1111}$ is the depolarization ratio of the resonant FWM field, $\chi^{(3)R}_{1221}$ and $\chi^{(3)R}_{1111}$ are components of the third-order nonlinear susceptibility, and $\boldsymbol{E_1}$ and $\boldsymbol{E_2}$ are the electric fields of NIR and UV lasers, respectively. In our experiment, taking the 204 nm radiation as an example, its intensity generated in the case of parallel polarization (i.e., $\boldsymbol{E_1} \parallel \boldsymbol{E_2}$) is about 3.3 times of that in the case of orthogonal polarization (i.e., $\boldsymbol{E_1} \perp \boldsymbol{E_2}$). From this measured value, we obtain $\bar{\rho} = 0.54$. For the case of $\boldsymbol{E_1} = E_1\hat{\boldsymbol{y}}$, $\boldsymbol{E_2} = \frac{E_2}{\sqrt{1+\xi^2}}(\hat{\boldsymbol{x}} + i\xi\hat{\boldsymbol{y}})$, Eq. (1) becomes

$$\boldsymbol{P_R} = \frac{3E_1^2 E_2}{\sqrt{1+\xi^2}}\chi^{(3)R}_{1111}(\omega_{\text{FWM}};\omega_1,\omega_1,\omega_2) \\ \times (0.54\hat{\boldsymbol{x}} + i\xi\hat{\boldsymbol{y}}). \quad (2)$$

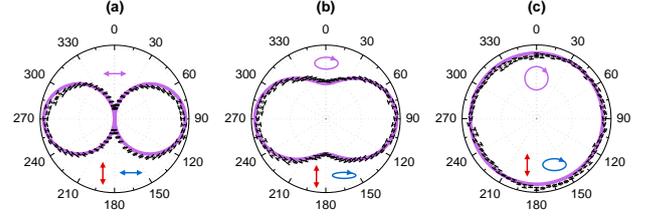

FIG. 3. (a) The nearly linearly polarized, (b) elliptically polarized, and (c) circularly polarized 204 nm radiation produced by the UV laser with $\xi$ =0.07, 0.27 and 0.52, respectively. The error bars represent the standard deviations of measurements. The solid lines represent the results calculated by Equation (2) with $\xi$=0, 0.27 and 0.54. The polarizations of the input NIR and UV pump lasers and the 204 nm radiation are denoted as directional arrows, ellipses or circles in insets (NIR: red; UV: blue; 204 nm: purple).

Eq. (2) clearly shows that we can generate arbitrarily polarized DUV/VUV coherent radiations by simply changing the ellipticity $\xi$ of UV laser. Experimentally, the main axis of the UV laser field is fixed in $\hat{\boldsymbol{x}}$ direction by using the combination of a half-wave plate and a quarter-wave plate. The polarization state of the 204 nm radiation is measured with a Rochon prism, and the polarization sensitivity of the prism and the spectrometer has been calibrated. Figure 3 demonstrates the polarization of the 204 nm radiation can be tuned from linear to circular. These experimental results are in good agreement with theoretical predictions of Eq. (2). The ellipticity of 204 nm radiation reaches about 0.9 when the UV laser with $\xi$=0.52 is employed (see Fig. 3(c)). The deviation between the theoretical and measured results in Fig. 3(a) is caused by the imperfect linear polarization of the UV laser field. Besides, the helicity of the 204 nm radiation is identical with that of the UV laser. The experimental results confirm that this scheme enables the control for polarization of DUV/VUV radiations with great flexibility in terms of ellipticity, helicity and main-axis direction.

As demonstrated above, DUV/VUV coherent radiations with narrow linewidth, selective wavelength and controllable polarization can be produced through the resonant interaction of femtosecond laser pulses with CO$^+$ ions. DUV/VUV Raman spectroscopy shows great advantages in the applications of biological identification [26] and explosives sensing [27], because the DUV/VUV excitation not only allows for the higher Raman scattering efficiency [28] but also effectively suppresses interference [29]. Thus, we try to show the application of our DUV/VUV light sources in Raman spectroscopy. It should be emphasized that the generated DUV/VUV radiation has an ideal spatial overlap with the femtosecond pump laser, and our scheme can uniquely combine the advantages of impulsive preparation of rotational coherences with the femtosecond pump laser and high spectral resolution with the DUV/VUV radiation. These inherent merits facilitate efficient Raman scattering to form a Raman comb in the DUV/VUV region. To verify the concept, the 204 nm radiation is chosen, and gaseous $O_2$ molecules are used as the Raman active medium due to the large Raman cross section and favorable rotational constant [30]. When the residual femtosecond laser exiting from the CO gas chamber travels through $O_2$ gas, the $O_2$ molecules are forced to rotate in phase as a result of impulsive alignment. Concerted rotation of molecules will efficiently modulate optical polarizability of the medium. The interaction of the 204 nm radiation with the modulated medium will give rise to high-order Raman scattering.

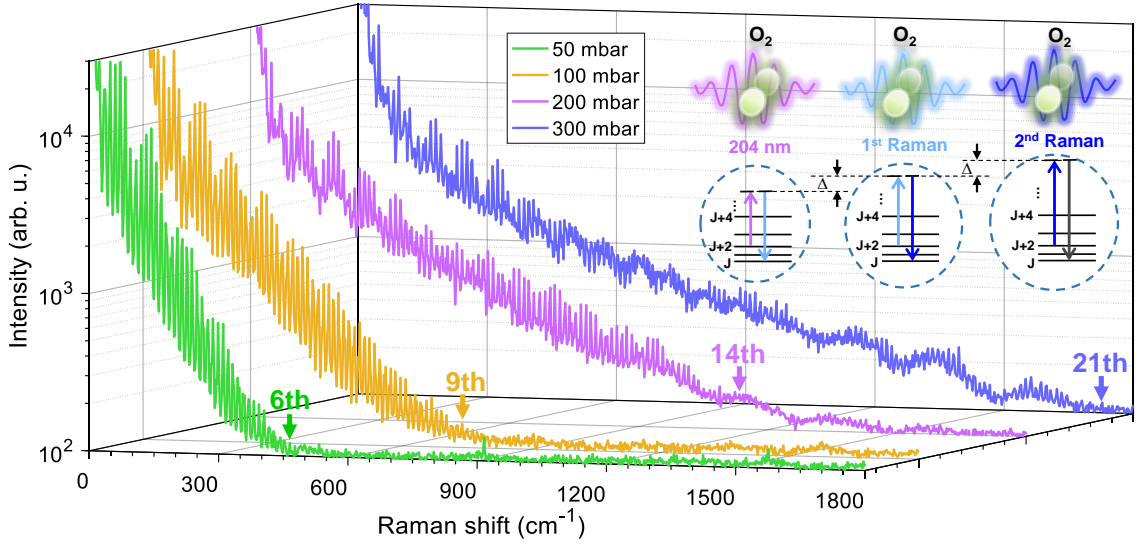

FIG. 4. Raman spectra recorded at different gas pressures. The arrows indicate the maximum orders of Raman scattering at different gas pressures. Inset is a schematic diagram for generation of high-order rotational Raman scattering, where the frequency shift Δ is equal to (4J+6)B.

Figure 4 shows anti-Stokes Raman spectra captured at different $O_2$ gas pressures, which consist of many discrete peaks. Their frequency shifts with respect to the 204 nm radiation perfectly match with the calculated values of high-order rotational Raman scattering. For the first order rotational Raman scattering, Raman shift obeys $\Delta=(4J+6)B$ with the rotational constant $B=1.4456$ cm$^{-1}$ of $O_2$ molecules [30]. Because only odd rotational states are populated for $O_2$ molecules, Raman sidebands consist of discrete peaks spaced by a constant value of 8B except for $J=0$ case. The Raman signal can be further scattered to produce higher order scattering, as conceptually illustrated in inset of Fig. 4. This cascading process is quite efficient due to the pre-prepared rotational coherence by the femtosecond pump laser. Two adjacent orders of Raman scattering are separated by 58B as observed in Fig. 4, which reflects the maximum population in $J=13$ state in the temperature of 294 K. One can see that Raman scattering is extended to higher orders with the increase of $O_2$ gas pressure. In the 300 mbar oxygen gas, we can observe Raman scattering up to 21th order, which forms a Raman comb with more than 100 sidebands. The Raman comb in the DUV/VUV region has a quasi-periodic spectral structure with the frequency spacing of 0.35 THz (i.e., 8Bc), and each Raman peak inherits the narrow linewidth of the 204 nm DUV radiation.

Owing to the coherence in the impulsively excited rotational wavepackets, the Raman sidebands are phase locked at alignment moments which can generate an ultrashort pulse train in the DUV/VUV regime. To confirm this, we performed a cross-correlation measurement to obtain the temporal information of these Raman signals. Figure 5(a) shows the difference-frequency signal of the Raman comb produced in 200-mbar $O_2$ gas together with the residual 204 nm radiation and the third 800 nm femtosecond laser beam as a function of their relative delay. The measured result clearly indicates that the 204 nm radiation is modulated at revival moments of coherently excited $O_2$ molecules. The spectral integration of anti-Stokes components in Fig. 5(a) reflects the temporal structure of the Raman comb. As demonstrated in Fig. 5(b), the Raman comb constructs a femtosecond pulse train with an interval of ~2.88 ps. The interval is equal to one quarter of rotational period $T_{rev}$ of $O_2$ molecules [30]. The relative intensity of these sub-pulses basically follows the temporal envelope of the 204 nm radiation. Furthermore, the characteristics of Raman comb such as the wavelength range and frequency spacing can be controlled using different DUV/VUV sources and different Raman-active molecules.

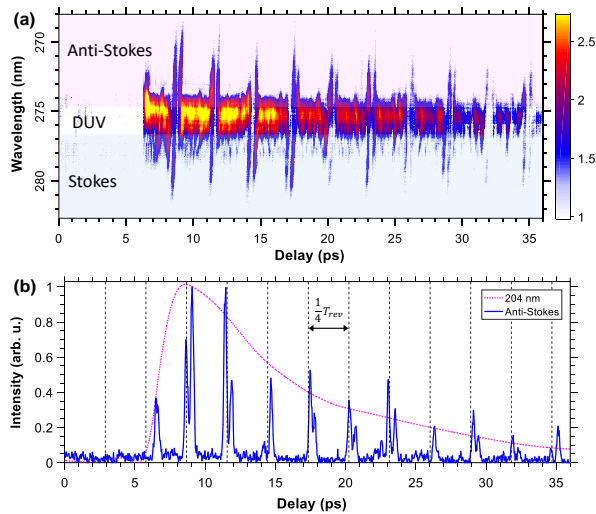

FIG. 5. (a) The time-frequency diagram obtained by the cross-correlation technique (logarithmic color scale). The time-frequency diagram is divided into three regions. The upper, middle and lower regions correspond to the temporal information of anti-Stokes Raman signal, the 204 nm radiation, and Stokes Raman signal, respectively. (b) Temporal structure of the Raman comb obtained by integrating over anti-Stokes spectral components in Fig. 5(a). Note that the first peak around 6.5 ps is not the Raman signal, because it still exists in argon gas. For comparison, the temporal envelop of the 204 nm radiation is indicated with a pink dash line.

To conclude, we have demonstrated generation, wavelength selection, polarization control and application of the DUV/VUV coherent radiations driven by coherence of strong-field ionized molecules. Specifically speaking, our scheme is based on resonant FWM in molecular ions. In principle, the scheme proposed in this work can be used for generating picosecond coherent sources in the exotic wavelength ranges such as VUV and infrared regimes. The abundant enengy levels of molecular ions provide a large number of operation wavelengths of such sources. These advantages will benefit a broad range of applications ranging from high-resolution spectroscopy and surface science to photochemistry and medicine, etc.

This work is supported by National Key R&D Program of China (2019YFA0705000); National Natural Science Foundation of China (11822410, 11734009); Strategic Priority Research Program of Chinese Academy of Sciences (XDB16030300); Key Research Program of Frontier Sciences of Chinese Academy of Sciences (QYZDJ-SSW-SLH010); Shanghai Municipal Science and Technology Major Project (2019SHZDZX01); Youth Innovation Promotion Association of CAS (2018284).

#Theses authors contributed equally to this work.
†jinpingmrg@163.com
*ya.cheng@siom.ac.cn